\documentstyle[]{mn}
\newif\ifAMStwofonts
\ifoldfss
  \newcommand{\rmn}[1] {{\rm #1}}

  \ifCUPmtlplainloaded \else
    \NewTextAlphabet{textbfit} {cmbxti10} {}
    \NewTextAlphabet{textbfss} {cmssbx10} {}
    \NewMathAlphabet{mathbfit} {cmbxti10} {} % for math mode
    \NewMathAlphabet{mathbfss} {cmssbx10} {} %  "   "    "
  \fi
  \ifAMStwofonts
    \ifCUPmtlplainloaded \else
      \NewSymbolFont{upmath} {eurm10}
      \NewSymbolFont{AMSa} {msam10}
      \NewMathSymbol{\upi}     {0}{upmath}{19}
      \NewMathSymbol{\umu}     {0}{upmath}{16}
      \NewMathSymbol{\upartial}{0}{upmath}{40}
      \NewMathSymbol{\leqslant}{3}{AMSa}{36}
      \NewMathSymbol{\geqslant}{3}{AMSa}{3E}

      \let\leq=\leqslant 
      \let\geq=\geqslant 
    \fi
  \fi
\fi % End of OFSS

\ifnfssone
  \newmathalphabet{\mathit}
  \addtoversion{normal}{\mathit}{cmr}{m}{it}
  \addtoversion{bold}{\mathit}{cmr}{bx}{it}
  \newcommand{\rmn}[1] {\mathrm{#1}}

  \newmathalphabet{\mathbfit} % math mode version of \textbfit{..}
  \addtoversion{normal}{\mathbfit}{cmr}{bx}{it}
  \addtoversion{bold}{\mathbfit}{cmr}{bx}{it}
  \newmathalphabet{\mathbfss} % math mode version of \textbfss{..}
  \addtoversion{normal}{\mathbfss}{cmss}{bx}{n}
  \addtoversion{bold}{\mathbfss}{cmss}{bx}{n}
  \ifAMStwofonts
    \ifCUPmtlplainloaded \else
      \UseAMStwoboldmath
      \makeatletter
      \new@mathgroup\upmath@group
      \define@mathgroup\mv@normal\upmath@group{eur}{m}{n}
      \define@mathgroup\mv@bold\upmath@group{eur}{b}{n}
      \edef\UPM{\hexnumber\upmath@group}
      \new@mathgroup\amsa@group
      \define@mathgroup\mv@normal\amsa@group{msa}{m}{n}
      \define@mathgroup\mv@bold\amsa@group{msa}{m}{n}
      \edef\AMSa{\hexnumber\amsa@group}
      \makeatother
      \mathchardef\upi="0\UPM19
      \mathchardef\umu="0\UPM16
      \mathchardef\upartial="0\UPM40
      \mathchardef\leqslant="3\AMSa36
      \mathchardef\geqslant="3\AMSa3E

      \let\leq=\leqslant 
      \let\geq=\geqslant 
    \fi
  \fi
\fi % End of NFSS release 1

\ifnfsstwo
  \newcommand{\rmn}[1] {\mathrm{#1}}

  \DeclareMathAlphabet{\mathbfit}{OT1}{cmr}{bx}{it}
  \SetMathAlphabet\mathbfit{bold}{OT1}{cmr}{bx}{it}
  \DeclareMathAlphabet{\mathbfss}{OT1}{cmss}{bx}{n}
  \SetMathAlphabet\mathbfss{bold}{OT1}{cmss}{bx}{n}
  \ifAMStwofonts
    \ifCUPmtlplainloaded \else
      \DeclareSymbolFont{UPM}{U}{eur}{m}{n}
      \SetSymbolFont{UPM}{bold}{U}{eur}{b}{n}
      \DeclareSymbolFont{AMSa}{U}{msa}{m}{n}
      \DeclareMathSymbol{\upi}{0}{UPM}{"19}
      \DeclareMathSymbol{\umu}{0}{UPM}{"16}
      \DeclareMathSymbol{\upartial}{0}{UPM}{"40}
      \DeclareMathSymbol{\leqslant}{3}{AMSa}{"36}
      \DeclareMathSymbol{\geqslant}{3}{AMSa}{"3E}

      \let\leq=\leqslant 
      \let\geq=\geqslant 
    \fi
  \fi
\fi % End of NFSS release 2

\ifCUPmtlplainloaded \else
  \ifAMStwofonts \else % If no AMS fonts
    \def\upi{\pi}
    \def\umu{\mu}
    \def\upartial{\partial}
  \fi
\fi

\newcommand{\samethanks}
	{{\Huge $^\star$}}

\newcommand{\eq}[1]
	{equation~(\ref{equation:#1})}

\newlength{\singlefigureheight}
\newlength{\doublefigureheight}
\newlength{\triplefigureheight}
\newlength{\squarefigureheight}
\newcommand{\FL}
        {FL}

\begin{document}

\title[Binary quasars]
	{Binary quasars}

\author[D.\ J.\ Mortlock, R.\ L.\ Webster and P.\ J.\ Francis]
        {
	Daniel J.~Mortlock,$^1$\thanks{
		E-mail:
        	dmortloc@physics.unimelb.edu.au (DJM);
		rwebster@physics.unimelb.edu.au (RLW);
		pfrancis@mso.anu.edu.au (PJF)}
	Rachel L. Webster$^{1}$\samethanks\
	and
	Paul J.~Francis$^{2,3}$\samethanks \\
        $^1$School of Physics, The University of Melbourne, Parkville,
        Victoria 3052, Australia \\
	$^2$Research School of Astronomy and Astrophysics,
		The Australian National University, Weston Creek,
		A.C.T.\ 2611, Australia \\
	$^3$Department of Physics and Theoretical Physics, Faculty of
                Science, The Australian National University, Canberra,
                A.C.T.\ 0200,\\ Australia \\
        }

\date{
Accepted 1999 June 15. 
Received 1999 June 15; in original form 1999 March 30}

\pagerange{\pageref{firstpage}--\pageref{lastpage}}
\pubyear{1999}

\label{firstpage}

\maketitle

\begin{abstract}
Quasar pairs are either 
physically distinct binary quasars or 
the result of gravitational lensing.
The majority of known pairs are in fact lenses, with a few confirmed as 
binaries, leaving a population of objects that have not yet been 
successfully classified.
Building on the arguments of Kochanek, Falco \& Mu\~{n}oz (1999a),
it is shown that there are no objective reasons to reject
the binary interpretation for most of these.
In particular, the
similarity of the spectra of the quasar pairs appears to be an artifact of 
the generic nature of quasar spectra.
The two ambiguous pairs discovered as part of the Large Bright
Quasar Survey (Q~1429$-$053 and Q~2153$-$0256) 
are analysed using principle components analysis, 
which shows that their spectral similarities are not greater than
expected for a randomly chosen pair of quasars from the survey.
The assumption of the binary hypothesis 
allows the dynamics, time-scales and separation distribution of binary quasars
to be investigated and constrained.
The most plausible model is that the quasars' activity is 
triggered by tidal interactions in a galatic merger,
but that the (re-)activation of the galactic nuclei occurs quite late
in the interaction, when the nuclei are within $80 \pm 30$ kpc 
of each other.
A simple dynamical friction model for the decaying orbits
reproduces the observed distribution of projected separations,
but the decay time inferred is comparable to a Hubble time.
Hence it is predicted that binary quasars are only observable 
as such in the early stages of galactic collisions, after which
the quiescent super-massive black holes orbit in the merger
remnant for some time.
\end{abstract}


\section{Introduction}
\label{section:introduction}

A small fraction ($\sim 1$ in 1000) of quasars are
observed to have at least one nearby quasar image with essentially the
same redshift 
(e.g.\ Kochanek 1995; Hewett et al.\ 1998),
and approximately 50 such systems are
known in all (e.g.\ Keeton \& Kochanek 1996).
Some of the pairs (and all of the higher multiples) are the result of 
gravitational lensing by intervening mass distributions;
some are physically distinct binary quasars.
In most cases the correct interpretation is quite clear,
even if considerable observational effort was required, but
there is a small number ($\sim 20$) of pairs of quasars which
cannot be categorised with certainty (e.g.\ Schneider 1993;
Kochanek, Falco \& Mu\~{n}oz 1999a; Peng et al.\ 1999).
These tend to have angular separations of $\ga 3$ arcsec,
and in most cases are treated as potential `wide separation'
lenses
(e.g.\ Narayan \& White 1988; Wambsganss et al.\ 1995; 
Mortlock, Webster \& Hewett 1996; Park \& Gott 1997; Williams 1997).
If they are lenses they
indicate the presence of a significant and otherwise unknown
population of dark objects with the mass of groups or clusters
of galaxies (e.g.\ Kochanek 1995; Hawkins 1997).
If they are binaries
they are almost certainly the result of enhanced fuelling 
in galactic mergers or interactions (Djorgovski 1991; 
Kochanek et al.\ 1999a).

The arguments in support of both possibilities
are summarised in Sec.~\ref{section:pairs}.
In Sec.~\ref{section:spectra} principal components analysis
is used to objectively test the spectral similarity of several
quasar pairs.
The binary hypothesis is subsequently adopted for the
entire pair population, and the implications
for the dynamics of binary quasars 
are explored in
Sec.~\ref{section:dynamics}.
The conclusions are summarised in
Sec.~\ref{section:conclusion}, together with a discussion of 
the future observational and theoretical possibilities.

\section{Known quasar pairs}
\label{section:pairs}

\begin{table*}
\begin{minipage}{173.5mm}
\caption{Possible binary quasars.}
\begin{tabular}{lccccrrrrrr}
\hline
Name & Status & Type & $\Delta m$ & $z$ 
	& \multicolumn{1}{c}{$|\Delta v_{||}|$} &
	\multicolumn{1}{c}{$\Delta \theta$}
        & & \multicolumn{1}{c}{$R$} & & \multicolumn{1}{c}{Ref.} \\
& & & & & &
	& \multicolumn{1}{c}{$\Omega_{\rmn m_0} = 1$} 
	& \multicolumn{1}{c}{$\!\! \Omega_{\rmn m_0} = 0.3 \!\!$}
	& \multicolumn{1}{c}{$\Omega_{\rmn m_0} = 0.3$} & \\
& & & & & &
	& \multicolumn{1}{c}{$\Omega_{\Lambda_0} = 0$} 
	& \multicolumn{1}{c}{$\!\! \Omega_{\Lambda_0} = 0 \!\!$}
        & \multicolumn{1}{c}{$\Omega_{\Lambda_0} = 0.7$} & \\
& & & & & & & & & & \\
MG~0023+171 & Binary? & $O^2R^2$ & 1.2 & 0.95 & $170\pm150$ km s$^{-1}$ 
	& 4\farcs8 & 29 kpc & 34 kpc & 38 kpc & 1\\
Q~0101.8$-$3012 & Unknown & $O^2$ & 0.8 & 0.89 & $0\pm200$ km s$^{-1}$ 
	& 17\farcs0 & 101 kpc & 117 kpc & 132 kpc & 2 \\
Q~0151+0448\footnote{Q~0151+0448 is also known as PHL~1222 and UM~144.}
        & Binary? & $O^2$ & 3.6 & 1.91 & $520\pm160$ km s$^{-1}$ & 3\farcs3
        & 19 kpc & 25 kpc & 28 kpc & 3 \\
QJ~0240$-$343 & Unknown & $O^2$ & 0.8 & 1.41 & $250\pm180$ km s$^{-1}$ 
	& 6\farcs1 & 37 kpc & 46 kpc & 52 kpc & 4 \\
Q~1120+0195\footnote{Q~1120+0195 is also known as UM~425.}
        & Unknown & $O^2$ & 4.6 & 1.47 & $200\pm100$ km s$^{-1}$ & 6\farcs5
        & 40 kpc & 49 kpc & 55 kpc & 5 \\
PKS~1145$-$071 & Binary & $O^2R$ & 0.8 & 1.45 & $80\pm60$ km s$^{-1}$ 
	& 4\farcs2
        & 26 kpc & 32 kpc & 35 kpc & 6 \\
Q~1208+1011 & Lens? & $O^2$ & 1.5 & 3.80 
	& $1200\pm1200$ km s$^{-1}$ & 0\farcs5
        & 2 kpc & 3 kpc & 3 kpc & 7 \\
HS~1216+5032 & Binary & $O^2R$ & 1.8 & 1.45 & $260\pm1000$ km s$^{-1}$ 
	& 9\farcs1 & 56 kpc & 69 kpc & 77 kpc & 8 \\
Q~1343+2640 & Binary & $O^2R$ & 0.1 & 2.03 
	& $200\pm900$ km s$^{-1}$ & 9\farcs5
        & 55 kpc & 72 kpc & 79 kpc & 9 \\
Q~1429$-$0053 & Unknown & $O^2$ & 3.1 & 2.08 & $260\pm300$ km s$^{-1}$ 
	& 5\farcs1 & 30 kpc & 39 kpc & 42 kpc & 10 \\
Q~1634+267 & Unknown & $O^2$ & 1.6 
	& 1.96 & $30\pm90$ km s$^{-1}$ & 3\farcs8
        & 22 kpc & 29 kpc & 32 kpc & 11 \\
J~1643+3156 & Binary & $O^2R$ & 0.6 
	& 0.59 & $80\pm10$ km s$^{-1}$ & 2\farcs3 
	& 12 kpc & 14 kpc & 16 kpc & 12 \\
Q~2138$-$431  
	& Unknown & $O^2$ & 1.2 & 1.64 & $0\pm100$ km s$^{-1}$ & 4\farcs5
        & 27 kpc & 34 kpc & 38 kpc & 13 \\
Q~2153$-$2056 & Binary? & $O^2$ & 3.4 & 1.85 & $1100\pm1500$ km s$^{-1}$ 
	& 7\farcs8 & 46 kpc & 60 kpc & 66 kpc & 14 \\
MGC~2214+3550 & Binary & $O^2R$ & 0.5 & 0.88 & $150\pm400$ km s$^{-1}$ 
	& 3\farcs0 & 18 kpc & 21 kpc & 23 kpc & 15 \\
Q~2345+007 & Lens? & $O^2$ & 1.5 
	& 2.15 & $480\pm500$ km s$^{-1}$ & 7\farcs3
        & 42 kpc & 56 kpc & 61 kpc & 16 \\
\hline
\label{table:binaries}
\end{tabular}

`Status' summarises the current evidence concerning the nature of each
pair (See Sec.~\ref{section:criteria} for the classification
criteria, and Mortlock (1999) for a summary of the evidence 
pertaining to each pair.);
confirmed lenses are not included.
`Type' summarises the detected optical and radio emission of 
the pair using the notation of Kochanek et al.\ (1999a), where
$O^2R^2$ denotes a pair in which both components are detected 
in the radio (and optical); $O^2R$ denotes a pair with only one 
radio-loud component (which is hence a binary quasar);
and $O^2$ denotes a pair with no radio emission at all.
$\Delta m$ is the magnitude difference of each pair (in the optical);
$z$ is the redshift;
$|\Delta v_{||}|$ is the observed line-of-sight velocity difference;
$\Delta \theta$ is the angular separation of the two components;
and
$R$ is the projected physical separation of the two
quasars, given for the three cosmological models
described in Sec.~\ref{section:cosmology},
with $H_0 = 70$ km s$^{-1}$ Mpc$^{-1}$ assumed throughout.

References: 
	1.\ Hewitt et al.\ (1987);
	2.\ Boyle et al.\ (1998);
	3.\ Meylan et al.\ (1990);
	4.\ Tinney (1995);
	5.\ Meylan \& Djorgovski (1988, 1989);
	6.\ Meylan et al.\ (1987) and Djorgovski et al.\ (1987);
	7.\ Magain et al.\ (1992) and Maoz et al.\ (1992);
	8.\ Hagen et al.\ (1996);
	9.\ Crampton et al.\ (1988);
	10.\ Hewett et al.\ (1989);
	11.\ Djorgovski \& Spinrad (1984);
	12.\ Brotherton et al.\ (1999);
	13.\ Hawkins et al.\ (1997);
	14.\ Hewett et al.\ (1998);
	15.\ Mu\~{n}oz et al.\ (1998);
	16.\ Weedman et al.\ (1982).
\end{minipage}
\end{table*}

Table~\ref{table:binaries} lists all known quasar pairs which 
have not been confirmed as gravitational lenses, 
in order of increasing right ascension. 
This list has been compiled mainly from previous lens candidate 
compilations 
(Turner 1988; Surdej 1990a,b; Surdej et al.\ 1991;
Surdej 1991; Schneider, Ehlers \& Falco 1992; Surdej \& Soucail 1993;
Schneider 1993; Keeton \& Kochanek 1996; Kochanek et al.\ 1999a),
but also includes more recent discoveries, such as 
Q~0101.8$-$3012 \cite{bo98} and the small separation pair 
Q~1208+1011 (Magain et al.\ 1992; Maoz et al.\ 1992).
The criteria by which these pairs 
are given their tentative classifications
are discussed in Sec.~\ref{section:criteria},
and a detailed summary of the status of each pair is given in 
Mortlock \shortcite{mo99}.
Some statistical arguments relating only to the overall properties 
of the population of pairs are presented in 
Sec.~\ref{section:population}.

\subsection{Classification criteria}
\label{section:criteria}

From the discovery of the first gravitational lens 
(Walsh, Carswell \& Weymann 1979)
and the subsequent interest in lensed quasars, it was soon 
realised that reasonably objective criteria must be established 
to determine whether pairs were lenses (e.g.\ Webster \& Fitchett 1986;
Schneider et al.\ 1992; Kochanek 1993; Kochanek et al.\ 1999a).

A pair can only be positively confirmed as a binary if 
the spectra of the images are vastly different, if only one image
is radio-loud (i.e.\ an `$O^2R$' pair, in the notation of 
Kochanek et al.\ 1999a),
or if the quasars' host are detected.
The only necessary condition a binary must satisfy is
that its components' spectra not be `too similar' to each other,
the meaning of which is investigated further in 
Secs.~\ref{section:population} and \ref{section:spectra}.

The main sufficient conditions for a pair to be identified as 
a lens are:
the presence of more 
than two images\footnote{It is possible that a physical triple or quadruple
quasar exists, but this seems very unlikely, as discussed in 
Sec.~\ref{section:activation}.};
the measurement of a time-delay between the images;
or the detection of a plausible deflector.
Naively, both a sufficient and necessary condition for a pair to be
a lens is that its components'
spectra are identical, 
as gravitational lensing is achromatic.
However the difficulty in assessing just how similar two 
spectra are, combined with the generic nature of quasar spectra,
means that it is difficult to prove a pair is not a binary on these
grounds, as discussed further in Sec.~\ref{section:spectra}.
Spectral similarity is not a necessary condition either --
the light from a lensed quasar travels along
different lines-of-sight to the observer, which can result in a
number of achromatic effects.
Dust along one line-of-sight can redden individual images
(e.g.\ Falco et al.\ 1999);
microlensing by stellar-mass objects 
can magnify the
continuum of one image relative to the emission lines;
intrinsic variability coupled with the time-delay can result in 
the simultaneous spectra of components differing; and even the fact that 
the light that makes up each image comes from slightly
different points in the quasar can result in different
spectra being observed\footnote{It is possible, for example, for
a lens to have broad absorption 
features in only one component if the absorbing `clouds' 
around a quasar are sufficiently small.
The physical separation
of photons in the two light paths would be only a few pc as they
passed through the clouds, but it is possible that the
broad line region consists of sheets of plasma only a few metres thick
\cite{bl91}, in which case the two lines-of-sight are effectively
uncorrelated.
This possibility is why Q~0151+0448 \cite{me90}
and Q~2153$-$2056 \cite{he98} have 
been classified more tentatively here than by Peng et al.\ \shortcite{pe99}.}.
Further, the images of a lensed quasar can even yield different
redshifts -- the two components of Q~0957+561
have redshifts which differ by $\Delta z = 0.0023 \pm 0.00017$ \cite{wi80}.

Clearly the ambiguous quasar pairs do not satisfy any of these
sufficient conditions, although some come close. 
The generic uncertain pair has no visible deflector,
but qualitatively similar spectra, neither of which allows
a definite classification.

\subsection{The population of pairs}
\label{section:population}

If the 11 ambiguous pairs in Table~\ref{table:binaries}
are predominantly lenses, the 
statistical properties of the sample should match those 
of the confirmed lenses, and likewise if they 
are mainly binaries. Their are some 
potential pit-falls to this approach (e.g.\ the definition of 
the samples; the presence of recently discovered pairs, such
as Q~0101.8$-$3012,
which may not require the existence of a particularly dark lens)
but it is a potentially powerful method.

\begin{figure}
\includegraphics{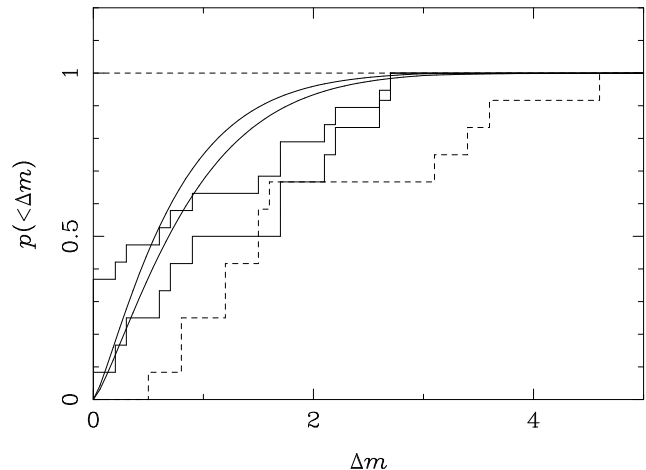}
\vspace{6.6cm}
\caption{A cumulative plot of the magnitude differences,
$\Delta m$, between
the possible lensed quasar pairs
(dashed line),
and all known two-image lenses, both with (upper solid line)
and without (lower solid line) the known arcs and rings (which are
generally treated as having $\Delta m = 0$).
The two smooth solid curves show the distribution predicted by a simple
lensing model for survey magnitude limits of $B = 19$ (upper line)
and $B = 21$ (lower line).}
\label{figure:delta_m}
\end{figure}

Kochanek et al.\ \shortcite{ko99a} used this technique to 
argue that most of the wide separation (i.e.\ $\Delta \theta 
\ga 3$ arcsec) pairs are not lenses for the following reasons.
The existence of the $O^2 R$ binaries, combined with 
the knowledge that most quasars are radio-quiet, implies 
a comparable or larger population of $O^2$ binaries -- 
presumably the majority of the 10 in Table~\ref{table:binaries}.
The distribution of the ambiguous pairs' redshifts 
is significantly different to that of the confirmed lenses, 
which is very difficult to account for in terms of lensing
(e.g.\ Williams 1997).
The distribution of flux ratios of the pairs and of the 
confirmed two-image lenses differ greatly
(Fig.~\ref{figure:delta_m}), although
neither are consistent with the distribution predicted
by a simple, spherical lens model, implying
that further investigation is warranted.
Lastly, if the pairs are lenses, the absence of a
smooth fall-off at larger separations is at odds with most 
theoretical models that predict wide separation lenses (e.g.\ 
Narayan \& White 1988; Kochanek 1995). 

The statistical objections to the binary hypothesis have
been the uncanny similarities of pairs' spectra 
(which is addressed in Sec.~\ref{section:spectra})
and the high number 
of pairs at such small separations
(relative to the predictions of the generic
quasar-quasar correlation function).
However,
Kochanek et al.\ \shortcite{ko99a} showed that the number of 
pairs observed was consistent with the number expected 
from the observed small-scale quasar-galaxy correlation function.
Further implications of this are explored in Sec.~\ref{section:dynamics}.

\section{Spectral similarities of quasar pairs}
\label{section:spectra}

The spectra of many of the pairs do appear unusually alike, but,
before this can be regarded as an objection to the binary 
hypothesis,
it must be shown quantitatively that they are significantly
more alike than those of randomly chosen quasars\footnote{Even
then, there are some reasons to think that binaries may have
similar spectra due to their common environment and formation time 
\cite{pe99}. However it is difficult to see how 
either 
could influence the localised velocity fields that are responsible 
for the emission line shapes.}.
There have been several attempts to implement this sort of test,
but the limitations of the data have proved critical in 
most cases. 
Turner et al.\ \shortcite{tu88b} found that the 
the emission line shapes of Q~1634+267 A and B were more
similar to each other than they were to those of another 
quasar of comparable luminosity and redshift, although the 
use of only a single comparison object limited the strength of 
the conclusions. 
Peng et al.\ \shortcite{pe99} found that the C\,{\sc iv} equivalent widths
and continuum slopes of 14 pairs (all those in Table~\ref{table:binaries}
bar Q~0101.8$-$3012 and Q~1208+1011) were more alike than 
over 97 per cent of comparable random quasar samples,
implying that the apparent spectral similarities are real. 
Lastly, Hawkins \shortcite{ha97} found that the 
colours of the two components of Q~2138$-$431 and Q~2345+007
were closer than expected, when compared 
to those of $\sim 10$ other quasars with similar redshifts. 
However the control quasars represent a heterogeneous sample,
which can only increase the {\em relative} similarity of the pairs' 
colours.
This illustrates why such relative tests are better at demonstrating
pairs can be binary objects. 
In a related investigation,
Small et al.\ \shortcite{sm97} showed that
the differences between the component spectra of
Q~1634+267 and Q~2345+007 were no larger than the temporal
variations in other quasars. Hence these pairs are consistent
with being lenses, despite the differences in their spectra.
However this represents only a necessary condition, and is not
a sufficient condition for the lensing interpretation.
The choice of experiment is probably
an artifact the greater interest in finding lenses.

The analysis that follows is directed towards determining
whether the two ambiguous quasar pairs in the 
Large Bright Quasar Survey (LBQS) are 
consistent with being binaries.
The data is described in  Sec.~\ref{section:lbqs}
and the method of comparison discussed in 
Sec.~\ref{section:pca}. The results and interpretation are
given in Sec.~\ref{section:results}.

\subsection{Quasar pairs in the Large Bright Quasar Survey}
\label{section:lbqs}

\begin{table*}
\begin{minipage}{138mm}
\caption{Quasar pairs in the LBQS.}
\begin{tabular}{lcccccrrr}
\hline
Name & Status & $m_{\rm A}$ & $m_{\rm B}$ & $\Delta m$ & $z$ 
	& \multicolumn{1}{c}{$|\Delta v_{||}|$} & $\Delta \theta$ 
	& \multicolumn{1}{c}{Ref.} \\
& & & & & & & & \\
Q~1009$-$0252 & Lens & $V = 17.9$ & $V = 20.5$ & 2.6 & 2.74
	& $150\pm200$ km s$^{-1}$ & 1\farcs5 & 1 \\
Q~1429$-$0053 & Unknown & $R = 17.7$ & $R = 20.8$ & 3.1 & 2.08 
	& $260\pm300$ km s$^{-1}$ & 5\farcs1 & 2 \\
Q~2153$-$2056 & Binary? & $B = 17.9$ & $B = 21.3$ & 3.4 & 1.85 
	& $1100\pm1500$ km s$^{-1}$ & 7\farcs8 & 3 \\
\hline
\label{table:lbqs}
\end{tabular}

`Status' summarises the current evidence concerning the nature of each
pair (See Table~\ref{table:binaries} and Sec.~\ref{section:criteria}.);
$m_{\rm A}$ and $m_{\rm B}$ are the magnitudes of the primary and
its companion, respectively, and $\Delta m$ is 
the difference between the two 
(The band is given in the table.);
$z$ is the redshift of the pair;
$|\Delta v_{||}|$ is the observed line-of-sight-velocity difference;
and
$\Delta \theta$ is the angular separation of the two components.

References: 
	1.\ Hewett et al.\ (1994);
	2.\ Hewett et al.\ (1989);
	3.\ Hewett et al.\ (1998).
\end{minipage}
\end{table*}

\begin{figure}
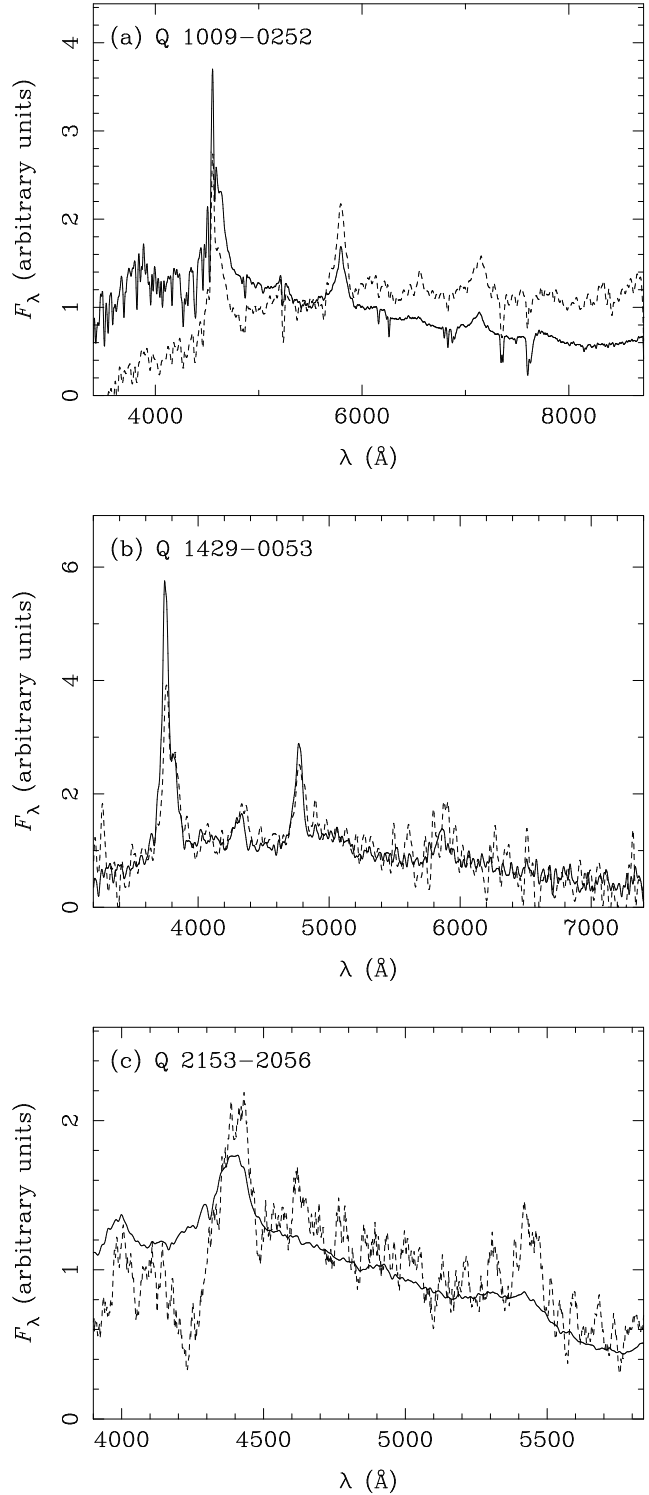

\includegraphics{1009.ps}
\includegraphics{1429.ps}
\includegraphics{2153.ps}
\vspace{20.02cm}
\caption{The spectra of the primaries (solid lines) and companions 
(dashed lines) of the three
quasar pairs in the LBQS: 
Q~1009$-$0252 (a);
Q~1429$-$0053 (b);
and 
Q~2153$-$2056 (c).
In each case, both spectra have been smoothed (with a smoothing length
between 5 and 10 \AA) and normalised to have an average flux of unity, 
to aid in the 
comparison. Note that all spectra are in the observed frame, not the
rest frame.}
\label{figure:spectra}
\end{figure}

The LBQS
is a sample of 1055 quasars with $m_{B_{\rmn J}} \ga 18.5$,
selected spectroscopically and without
any reference to morphology (Hewett, Foltz \& Chaffee 1995).
Due to both its size and its well-characterised
selection effects, it is an excellent source of data for
studying the similarities and differences between quasar spectra.

As part of the survey, there was a systematic search for companions
within $\sim 10$ arcsec of each quasar.
The companion search has sufficient
dynamic range ($\Delta m \simeq 3.5$)
to easily pick out most secondary lensed images, and also
to detect potentially dimmer binary companions.
The search has yielded three pairs so far
(listed in Table~\ref{table:lbqs}), and it is 
`unlikely that further pairs will be identified' \cite{he98}.
Q~1009$-$0252 \cite{he94} is a confirmed lens;
Q~1429$-$0053 \cite{he89} is a reasonable lens candidate;
and
Q~2153$-$2056 \cite{he98} is probably a binary quasar.
The spectra of all three pairs, taken concurrently in each case,
are shown in Fig.~\ref{figure:spectra},
and it is clear that there are significant differences between
the primary and companion of both Q~1009$-$0252 (the lens)
and Q~2153$-$2056.
However, a more quantitative, and hence objective, means 
of testing the similarity of the spectra is needed to proceed 
further.

\subsection{Comparison of quasar spectra}
\label{section:pca}

The most common method of analysing spectra (aside from 
visual inspection) is to measure the properties of 
individual features, such as emission lines and continuum
properties. The weakness of this approach is that
the classification of the features is usually subjective
at some level. 

It is preferable to use a completely 
quantitative method, such as 
principal components analysis (PCA; Whitney 1983;
Murtagh \& Hecht 1987; Mittaz, Penston \& Snijders 1990).
PCA can be used on pre-determined features, but, for the reasons 
given above, it is more powerful to perform PCA 
on the raw spectra.
This `spectral' PCA
has the advantages that it is a 
completely objective analysis, uses all the available data,
and is adept at dealing with low signal-to-noise
ratio spectra. 
With each spectrum treated as a vector in a multi-dimensional space,
the sample of spectra represents a distribution in this space,
the centroid of which is the mean spectrum.
Subtracting the mean from each spectrum yields a set of points 
(centred on the origin) which represents the variation between spectra.
The vector (or spectrum, of sorts) along which this 
distribution is most elongated is 
the first 
principle component (PC) of the sample. Subtracting it
from all the already mean-subtracted
spectra allows the procedure to be repeated, generating 
the subsequent components. 
The subtraction of only the first few PCs
usually (i.e.\ if the PCA is successful)
leaves a condensed, spherical distribution of 
points that is essentially a hyper-sphere of noise.
Almost all the information is contained the in the first few
components; the data 
compression involved can be large. Francis et al.\ \shortcite{fr92}
found that the LBQS spectra could be expressed in terms of 
coefficients of $\sim 10$
components, as opposed to several hundred wavelength bins.

Following Francis et al.\ \shortcite{fr92}, two subsets were
taken from the LBQS for the PCA.
Sample 1
contains 325 quasars (including all three pairs)
with $1.4 \leq z \leq 2.7$,
and rest-frame spectra
covering the range from 1400~\AA\ to 2200~\AA.
Importantly, this range covers several prominent
emission lines, notably C\,{\sc iv} (1549~\AA)
and 
Al\,{\sc iii}/C\,{\sc iii]} (1858~\AA/1909~\AA).
Sample 2
has only 209 quasars (including the the lens and 
the wide separation pair Q~1429$-$0053), but,
with $1.8 \leq z \leq 3.3$, covers a sligtly bluer part of rest-frame
spectrum: 1180~\AA\ to 1780~\AA. It includes the C\,{\sc iv}
emission line and the Ly\,$\alpha$/N\,{\sc v} complex
(1216~\AA/1240~\AA).
The other wide separation
pair, Q~2153$-$2056, could not be included in sample 2, as the 
available
spectra (Fig.~\ref{figure:spectra} (c)) do not include the 
Ly\,$\alpha$/N\,{\sc v} blend.
A number of quasars, including the
stated pairs, appear in both samples. 

\begin{figure}
\includegraphics{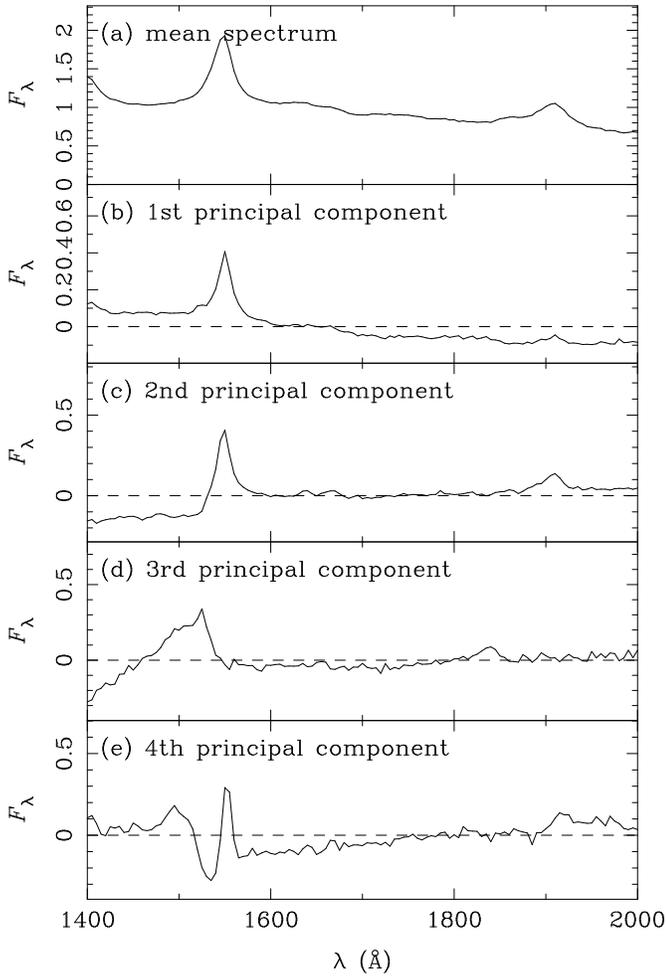}
\vspace{13.3cm}
\caption{The mean spectrum (a) and the first four
component `spectra' (b-e) of sample 1, plotted as a
function of rest-frame (i.e.\
in the frame of the quasars) wavelength.
The mean spectrum is normalised
to an average flux of unity,
and all the principal components have zero mean. The 3rd and 4th PCs
have been inverted so that emission features are positive.
The prominent emission lines are C\,{\sc iv} (at $\sim 1550$ \AA)
and the Al\,{\sc iii}/C\,{\sc iii]} blend (at $\sim 1900$ \AA).}
\label{figure:pcs_1}
\end{figure}

\begin{figure}
\includegraphics{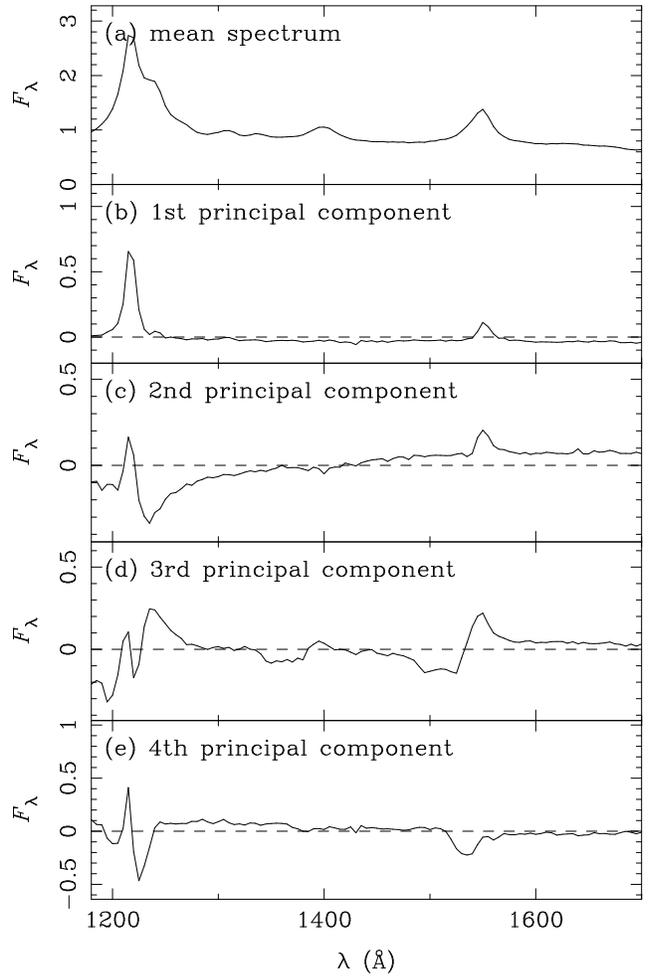}
\vspace{13.3cm}
\caption{The mean spectrum (a) and the first four
component `spectra' (b-e) of sample 2,
plotted as a function of rest-frame (i.e.\
in the frame of the quasars) wavelength.
The mean spectrum is normalised
to an average flux of unity,
and all the principal components have zero mean. The 3rd and 4th PCs
have been inverted so that emission features are positive.
The prominent emission lines are the Ly\,$\alpha$/N\,{\sc v}
blend (at $\sim 1230$ \AA) and C\,{\sc iv} (at $\sim 1550$ \AA).}
\label{figure:pcs_2}
\end{figure}

Spectral PCA was performed on both subsamples,
generating a set of PC spectra 
for each sample, and a set of coefficients for each quasar.
As expected, most of 
the variation between spectra was contained in the first few
components, which are shown in Fig.~\ref{figure:pcs_1} (for sample 1),
and Fig.~\ref{figure:pcs_2} (for sample 2).
The components are similar to those shown in Francis et al.\
\shortcite{fr92} (as expected, given the related set of input spectra), 
and their interpretation is also similar.
The first three components of both samples are related to:
the cores of the emission lines 
(PC 1; Figs.~\ref{figure:pcs_1} and \ref{figure:pcs_2} (b));
the continuum slope 
(PC 2; Figs.~\ref{figure:pcs_1} and \ref{figure:pcs_2} (c));
and 
broad absorption lines 
(PC 3; Figs.~\ref{figure:pcs_1} and \ref{figure:pcs_2} (d)).
The 4th PC of samples 1 and 2 relate to the wings of 
of the C\,{\sc iv} and 
Ly\,$\alpha$ emission lines, respectively 
(Figs.~\ref{figure:pcs_1} and \ref{figure:pcs_2} (e)). 
It is promising that the C\,{\sc iv} line is prominent in the first PC, 
as it is the most prominent difference between the spectra of several
of the confirmed binaries (Q~1216+5032 and Q~1343+2640), 
and both Turner et al.\ \shortcite{tu88} and Peng et al.\ 
\shortcite{pe99} found its properties revealing.

Whilst the PCA has enabled
each spectrum to be specified in terms of several numbers
(i.e.\ the coefficients of the first few PCs), 
each spectrum is still a multi-dimensional vector. 
The data can be further compressed by defining a metric in the
PC space.
The PCA naturally scales 
the variance in each component with its relative importance,
so a simple Euclidean metric is chosen.
The difference, $D$, between the spectra
of two quasars A and B is then given by
\begin{equation}
\label{equation:D}
D_{\rmn AB} = \left[ \sum_{n = 1}^{n_{\rmn max}} 
(C_{{\rmn A},n} - C_{{\rmn B},n})^2 \right]^{1/2},
\end{equation}
where $C_{{\rmn A},n}$ and $C_{{\rmn B},n}$ are the 
coefficients of the $n$th 
principal component of spectra of A and B, respectively, and $n_{\rmn max}$ is
the highest component used. The latter does not have an appreciable 
affect on the results, as long as the first few components are included,
so $n_{\rmn max} = 4$ was used.

\begin{figure}
\includegraphics{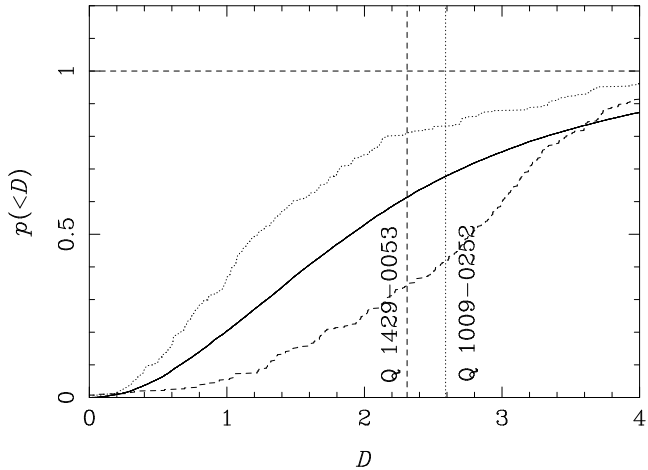}
\vspace{6.6cm}
\caption{The cumulative distribution of the difference
between two spectra, $D$, shown for sample 2 (Sec.~\ref{section:pca}).
The solid line is the distribution derived from all possible
pairs of spectra in the sample; the other lines are from the
differences between each of the primaries of the two pairs
(Q~1429$-$0053: dashed line;
Q~1009$-$0252: dotted line)
and the rest of the sample. The two vertical lines show
difference between each primary and its respective companion.}
\label{figure:ks_pc_2}
\end{figure}

Two sets of $D$-values were created for each sample: 
the differences between all the spectra in the sample;
and the differences between each 
of the pairs' primaries and the rest of the sample
(which determines the fraction of spectra
in the sample that are closer to the primary in question than
its companion).
The resultant cumulative plots are shown for sample 1 
and sample 2 in Figs.~\ref{figure:ks_pc_1} and \ref{figure:ks_pc_2},
respectively.

\begin{figure}
\includegraphics{ks_pc_1.ps}
\vspace{6.6cm}
\caption{The cumulative distribution of the difference
between two spectra, $D$, shown for sample 1 (Sec.~\ref{section:pca}).
The solid line is the distribution derived from all possible
pairs of spectra in the sample; the other lines are from the
differences between each of the primaries of the three pairs
(Q~1429$-$0053: dashed line;
Q~2153$-$2056: dot-dashed line;
Q~1009$-$0252: dotted line)
and the rest of the sample. The three vertical lines show
difference between each primary and its respective companion.}
\label{figure:ks_pc_1}
\end{figure}

\subsection{Implications for the quasar pairs}
\label{section:results}

The two components of the confirmed lens, Q~1009$-$0252, are clearly 
different. This is not unexpected -- as discussed in 
Sec.~\ref{section:criteria}, there are a number of reasons 
why lensed spectra need not be the same.
However, this analysis shows that they can actually be less similar
than a two spectra chosen at random: Fig.~\ref{figure:ks_pc_1} 
shows that 75 per cent of random 
pairs are more alike than the components of the lens.
However the main difference between the two spectra of the
Q~1009$-$0252 is the
reddening of the companion spectrum (Fig.~\ref{figure:spectra} (a)),
which is probably
due to dust along the line-of-sight to the fainter image
(see Sec.~\ref{section:criteria} and Falco et al.\ 1999).

The components of Q~2153$-$2056 are also more dissimilar than 
those of random pairs.
Naively, the lesson of Q~1009$-$0252
implies that Q~2153$-$2056 cannot be rejected 
as a lens on this basis alone.
The difference between the spectra of Q~2153$-$2056 cannot 
be so readily explained in terms of lensing -- indeed it 
appears that the companion is a broad absorption line quasar,
which is only explained 
simply\footnote{See Sec.~\ref{section:criteria} for a complex explanation.}
if they are two distinct objects.
This distinction reveals one of the limitations of PCA -- 
there is no physics or plausibility input, and so it cannot tell
that some spectral differences (specifically Q~1009$-$0252) 
might be due to dust. In this instance, a continuum fit to 
the spectra would circumvent this problem.
Due to both the spectral differences and the absence of a visible
deflector, Hewett et al.\ \shortcite{he98} interpret Q~2153$-$2056 as 
a probable binary, a conclusion which is confirmed here.

The results for Q~1429$-$0053 (which is a more probable lens a priori)
are more interesting.
For the wavelength range covered by sample 1, 
the components of Q~1429$-$0053 are more similar than
$\sim 85$ per cent of random pairs. It also appears that 
Q~1429$-$0053 is a slightly unusual quasar spectrally, and so
only $\sim 9$ per cent of the spectra are more similar to 
the primary than its companion. 
The sample 2 results paint a slightly different
picture. Q~1429$-$0053 B is no more similar to Q~1429$-$0053 A
than a random quasar from the sample. This is despite the
fact that Q~1429$-$0053 A is at least slightly unusual -- 
the dashed line and the solid line (representing the whole
sample) are quite separated in Fig.~\ref{figure:ks_pc_2}.
Such results are perfectly consistent with Q~1429$-$0053 being
a lens (c.f.\ Q~1009$-$0252), but are also 
perfectly consistent with the two spectra having been drawn at random 
from the LBQS, and hence consistent with Q~1429$-$0053 being
two distinct quasars. 
However, 
almost all the difference between
the two components of Q~1429$-$0053 can
be attributed to the relative strengths of the Ly\,$\alpha$-N\,{\sc v}
complex (appearing at $\sim 4800$ \AA\ in Fig.~\ref{figure:spectra} (b)),
the equivalent widths differ by
$\sim 50$ per cent. 
The possibility of microlensing of the 
the secondary image (See Sec.~\ref{section:criteria}.) means that 
the lens interpretation is still possible, although
the macrolensing flux ratio this would imply
($\Delta m \simeq 3.5$) is 
very unlikely a priori (e.g.\ Fig.~\ref{figure:delta_m}). 

There are several potential systematic errors which,
whilst they do not affect the above conclusions, 
are important in principle.
The samples covered a wide redshift range, 
possibly increasing the spread in the samples' spectra;
this could result in a true binary being rejected,
but could not result in a lens being wrongly classed a binary.
Another potential bias that is unimportant is the faintness of the
companion images in these pairs -- lower luminosity quasars
might have different spectra. 
If the pairs are lenses this is clearly irrelevant,
as they are not intrinsically so faint. 
If they are binaries, any bias that might make the spectra 
appear more dissimilar is acceptable, even desirable.

The main source of random uncertainty is the low quality
of the companions' spectra (despite the fact that 
PCA is adept at analysing low
signal-to-noise data), which is simply a function of 
their faintness (Table~\ref{table:lbqs}). 
Several hours' observation on a 10-m class telescope
would yield companion spectra of better quality than the current
primary spectra. 

On the basis of these results and the existing work summarised in 
Sec.~\ref{section:pairs}, all the ambiguous pairs are assumed 
to be binaries in Sec.~\ref{section:dynamics}. 
This, however, is quite an adventurous extrapolation -- 
the broader but less rigorous results of Peng et al.\ \shortcite{pe99}
imply that the spectra of the pair population are unusually alike.

\section{Dynamics of binary quasars}
\label{section:dynamics}

If all the quasar pairs listed in Table~\ref{table:binaries} are physical 
binaries, there are far too many to represent a 
simple a small-scale manifestation of quasar 
clustering \cite{dj91}.
The fairly natural conclusion (Djorgovski 1991; Schneider 1993)
is that any close quasar pairs are the
result of increased nuclear activity in interacting galaxies.
Kochanek et al.\ \shortcite{ko99a} showed that the 
number of binaries matches the number expected 
from the quasar-galaxy correlation function at small scales.
Henceforth, all the pairs in Table~\ref{table:binaries} are
treated as physical binaries, and the merger model is 
also adopted.
The sample of pairs can then be used to constrain
the distribution of physical separations
(Sec.~\ref{section:orbits}), and also to investigate 
the formation and evolution of 
binary quasars (Sec.~\ref{section:physics}).

\subsection{Physical separations}
\label{section:orbits}

The projected separations of the quasar pairs can be 
inferred for a given cosmological model (Sec.~\ref{section:cosmology}),
but some method of inversion is required to obtain the
distribution of physical separations. 
A random deprojection is explored in 
Sec.~\ref{section:deproj} and 
a simple orbital model is discussed in 
Sec.~\ref{section:circles}.

\subsubsection{Cosmological models}
\label{section:cosmology}

The projected separations of the pairs are given by
$R = \Delta \theta\,d_{\rmn A} (0, z)$, where $d_{\rmn A}$ is 
the angular diameter distance, and so $R$ varies with the cosmological
model (and also within a model, due to weak lensing). 
Three models
are used: the Einstein-de Sitter (EdS) model
($\Omega_{\rmn m_0} = 1$ and $\Omega_{\Lambda_0} = 0$);
a low-density model
($\Omega_{\rmn m_0} = 0.3$ and $\Omega_{\Lambda_0} = 0$);
and a low-density, spatially-flat model
($\Omega_{\rmn m_0} = 0.3$ and $\Omega_{\Lambda_0} = 0.7$),
where
$\Omega_{\rmn m_0}$ is the ratio of the current density of the universe
to the critical density and $\Omega_{\Lambda_0}$ is the similarly
normalised cosmological constant (e.g.\ Carroll, Press \& Turner 1992).
The $\sim 10$ per cent error in Hubble's constant (taken to be
$H_0 = 70$ km s$^{-1}$ Mpc$^{-1}$ here) does 
not contribute significantly to the uncertainty in $R$.

\subsubsection{Random deprojection}
\label{section:deproj}

\begin{figure*}
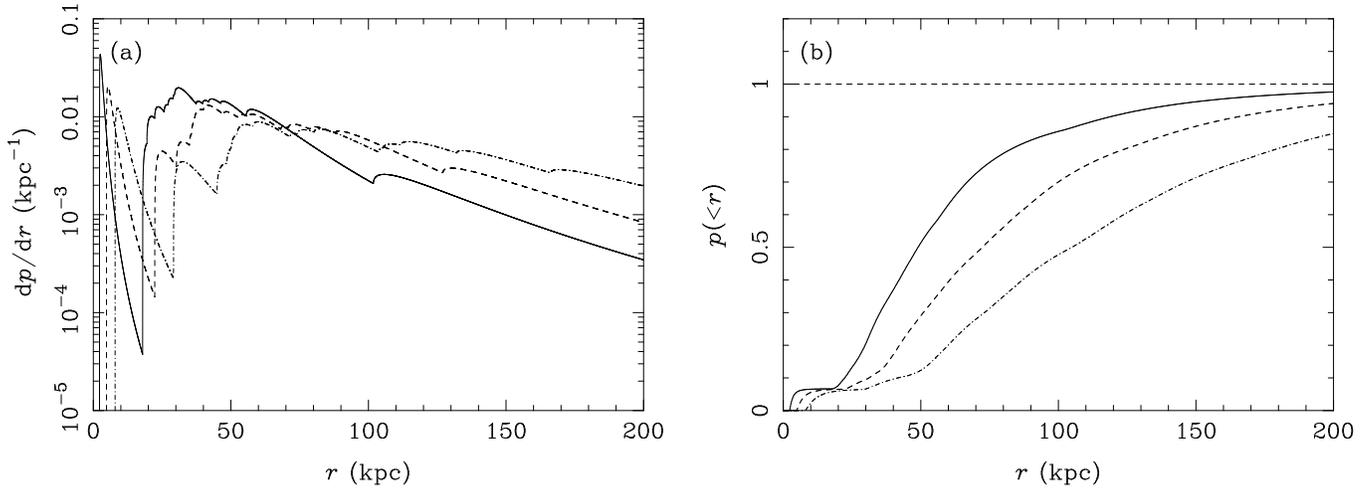

\includegraphics{dpdr_obs.ps}
\includegraphics{p_r_obs.ps}
\vspace{6.7cm}
\caption{The differential (a) and cumulative (b)
separation distribution of
binary quasars obtained by applying the random
deprojection described in Sec.~\ref{section:deproj}
to the data listed in Table~\ref{table:binaries}.
The three lines show the results assuming the three cosmological models
described in Sec.~\ref{section:cosmology}: the EdS model (solid line);
the low density model (dashed line) and the low density, flat
model (dot-dashed line).}
\label{figure:dpdr_obs}
\end{figure*}

The problem of deprojection -- effectively obtaining three dimensional
information from a two-dimensional image of the sky -- is an old one,
with a number of different partial solutions (e.g.\
Courant \& Hilbert 1962; Gerhard \& Binney 1996).
The problem at hand is 
to obtain a probability 
distribution of the physical separation, ${\rmn d}p/{\rmn d}r|_R$,
given an observed projected separation, $R$.
A Bayesian approach is 
adopted whereby the separation vector, $\bmath{r}$, 
is considered to be randomly-oriented,
a posteriori. 
Mathematically, 
${\rmn d}p/{\rmn d}i = \sin (i)$,
where $i$ is the (inclination) angle between $\bmath{r}$
and the line-of-sight. It is normalised for
the range $0 \leq i < \pi / 2$, and the projected separation is given
by $R = r | \sin (i) |$. From this, a change of variables
yields
\begin{eqnarray}
\left. \frac{{\rmn d} p}{{\rmn d} r} \right|_{R}
& = & \left| \frac{{\rmn d} p}{{\rmn d} i} \right|
        \left| \frac{{\rmn d} i}{{\rmn d} r} \right| \nonumber \\
& = & \left\{
        \begin{array}{lll}
        0, & {\rm if} & r < R, \\
        & & \\
        \frac{16}{\pi} \frac{R^3}{r^4}
        \sqrt{1 - \frac{R^2}{r^2}}, & {\rmn if}
        & r \geq R,
        \end{array}
        \right.
\label{equation:dpdr}
\end{eqnarray}
where the final distribution is also normalised to unity,
and has an expectation value of $16 / (3 \pi) R$, 
assuming $R > 0$.

Convolving ${\rmn d}p/{\rmn d}r|_R$ with the data gives an
approximation to the 
physical separation distribution of the binary quasars,
the differential and cumulative forms of which in Fig.~\ref{figure:dpdr_obs}.
This `random deprojection' shows a tail of pairs with quite large separations
and a possible 
paucity of small separation pairs. (The one pair with $r \la 10$ kpc,
Q~1208+1011 is a probable lens; if it were confirmed as such, the
small separation `hole' would be quite clear.)
Any hole cannot
explained by selection effects, as there are a large number of 
lenses with angular separations of $\sim 1$ arcsec, 
implying that any such binaries would have been found.
Kochanek et al.\ \shortcite{ko99a} showed that dynamical friction 
(Sec.~\ref{section:df}) can account for the
hole, although the
inferred separation distribution (linearly rising with $r$) is otherwise 
quite different from the 
approximately exponential fall-off apparent in Fig.~\ref{figure:dpdr_obs}.
Some clues as to the nature of small separation binaries might
be given by the discovery of 
three possible binaries in a sample of $\sim 100$ BL Lacs
\cite{sc99}.
The frequency of these pairs is far too high to be explained in
terms of gravitational lensing by galaxies, but the binary
interpretation is no more comfortable. It would require
not only that a few per cent of BL Lacs are binaries,
but that the optical jets of the companion BL Lacs are parallel.

\subsubsection{Orbital models}
\label{section:circles}

\begin{figure}
\includegraphics{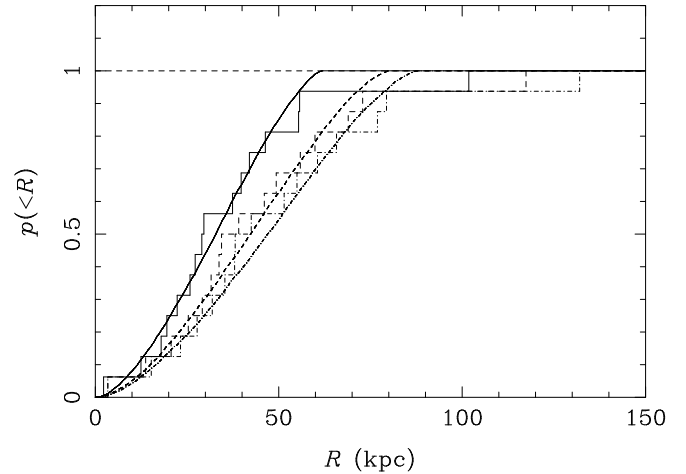}
\vspace{6.7cm}
\caption{The cumulative distribution of projected separation, $R$,
for the quasar pairs in Table~\ref{table:binaries}
and the best-fit dynamical friction model described in 
Sec.~\ref{section:circles}.
The three lines show the results assuming the three cosmological models
described in Sec.~\ref{section:cosmology}: the EdS model (solid line);
the low density model (dashed line) and the low density, flat
model (dot-dashed line).}
\label{figure:p_r_proj_fit}
\end{figure}

An approach almost opposite to the random deprojection used in
Sec.~\ref{section:deproj} is to assume an orbital
model for the binaries, and fit the resultant distribution of
projected separations to the data.
The paths of two quasars in a pair of merging galaxies is 
undoubtedly complex, but a natural first approximation is to 
assume that they are in decaying elliptical orbits.
However averaging over the random orientation
removes most of the effects of ellipticity, so circular orbits 
can be used in general \cite{mo99}.
The main difference was in the radius of closest approach, 
which may be relevant to the
activation of the quasars (Sec.~\ref{section:activation}).
The existence of the outer cut-off of between 50 kpc and 100 kpc is 
confirmed, independent of the distribution of separations used. 
The putative inner cut-off is less well constrained,
and, assuming ${\rmn d}p/{\rmn d}r \propto r$ 
(based on the dynamical friction approximation discussed in
Sec.~\ref{section:df}),
the data is consistent with a population of pairs extending
to zero separation, as shown in Fig.~\ref{figure:p_r_proj_fit}. 
However, formally better fits are obtained with
an inner cut-off of $\sim 10$ kpc, especially if 
the smallest separation pair, Q~1208+1011,
is a lens as suspected.
An inner cut-off would imply that
the 
at least one of the quasars in each pair `turns off' prior to the complete
decay of the orbits.
Possible reasons for this are discussed in Sec.~\ref{section:df}.

\subsection{Physical processes}
\label{section:physics}

If binary quasars are the result of galactic collisions 
during which {\em both} nuclei become active, 
the assumption that the systems are bound implies a minimum
mass for the systems
(Sec.~\ref{section:energy}),
and the distribution of physical separations 
(along with timing arguments)
can be used to investigate both the activation 
and evolution of the pairs
(Secs.~\ref{section:activation} and \ref{section:df},
respectively).

\subsubsection{Total energy}
\label{section:energy}

\begin{figure}
\includegraphics{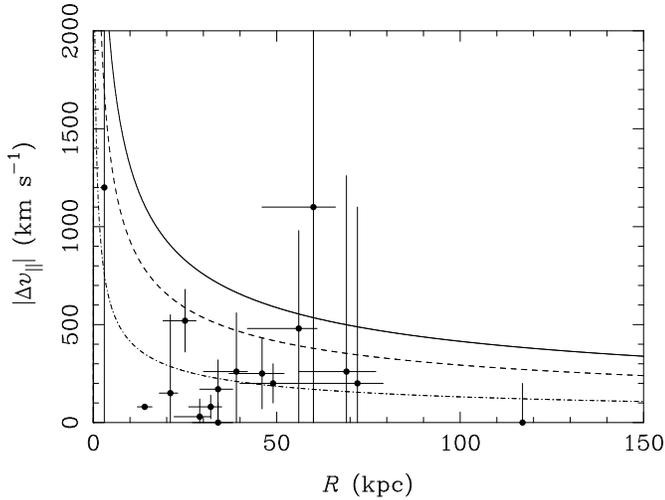}
\vspace{6.6cm}
\caption{A scatter plot showing the projected separation,
$R$, and the line-of-sight velocity difference, 
$\Delta v_{||}$, of the candidate binary quasars listed in 
Table~\ref{table:binaries}. The velocity error bars are random
measurement errors, but the separation errors are taken from the
spread in $R$ over cosmological models, and are thus
both random and 
systematic. Also shown is $E_{\rmn min} = 0$ [as defined in
\eq{e_min}], if {\em each} quasar's host 
galaxy weighed 
$10^{12} M_{\sun}$ (solid line),
$5 \times 10^{11} M_{\sun}$ (dashed line)
or
$10^{11} M_{\sun}$ (dot-dashed line).
If the binaries are bound, the points should lie
below the relevant line.}
\label{figure:e_min}
\end{figure}

It is possible that quasars could form in `glancing' collisions
(e.g.\ Noguchi 1987, 1988), but the absence of 
`post-collisional' pairs 
with larger separations argues against this for the binary 
population, implying that they exist in
gravitationally bound systems.
The minimum possible centre-of-mass frame energy of a pair
(obtained by setting the line-of-sight separation and projected velocity
difference to zero) is
\begin{equation}
\label{equation:e_min}
E_{\rmn min} = \frac{M_{\rmn A} M_{\rmn B}}{M_{\rmn A} + M_{\rmn B}}
	\frac{\Delta v^2_{||}}{2}
	- \frac{G M_{\rmn A} M_{\rmn B}}{R},
\end{equation}
where $G$ is Newton's gravitational constant, and $M_{\rmn A}$
and $M_{\rmn B}$ are the masses associated with the two quasars. 
Inverting \eq{e_min} and assuming $E_{\rmn min} = 0$
gives the hard lower limit for the mass of a bound system as 
\begin{equation}
M_{\rmn tot} = M_{\rmn A} + M_{\rmn B} 
\geq \frac{R \, \Delta v^2_{||}}{2 G}.
\label{equation:m_tot}
\end{equation}

Fig.~\ref{figure:e_min} shows a scatter plot of the 
projected separations and velocity differences of all the
pairs in Table~\ref{table:binaries}, with the large 
uncertainties in both $R$ and $\Delta v_{||}$ 
clearly important.
Despite this, 
$M_{\rmn tot} \ga 5 \times 10^{11} M_{\sun}$ is inferred,
confirming that quasars are typically in larger 
galaxies (Hooper, Impey \& Foltz 1997). However
using the random
deprojection disussed in Sec.~\ref{section:deproj}
to account for the projection effects 
yields a larger (but softer) limit of 
$M_{\rmn tot} \ga 2 \times 10^{12} M_{\sun}$
\cite{mo99}.

\subsubsection{(Re-)activation}
\label{section:activation}

It is widely believed that quasars are the
active nuclei of galaxies, where gas (and other material)
accretes onto a massive central black hole
(e.g.\ Lynden-Bell 1969; Rees 1984).
Many galaxies have massive but quiescent black holes at their centres
(e.g.\ Kormendy et al.\ 1996, 1997), and there are several plausible
explanations for how they may begin accreting and become active
galactic nuclei.
One important mechanism appears to be encounters between galaxies,
which can result in the formation of a massive black hole
(e.g.\ Rees 1984) or provide fuel to previously 
quiescent nuclei.
This is supported by {\em Hubble Space Telescope}
({\em HST}) observations which show that many 
quasars' host galaxies 
are distorted (Bahcall, Kirhakos \& Schneider 1995a;
Boyce, Disney \& Bleaken 1999)
and that low-redshift quasars are
often found to be physically associated with other galaxies (e.g.\
Yee 1987;
Bahcall, Kirhakos \& Schneider 1995b; Bahcall et al.\ 1997)
or involved in collisions \cite{st90}.
$N$-body simulations also show that gravitational torques
between colliding galaxies can cause gas to flow to their cores,
potentially acting as fuel for the central black holes \cite{ba96}.
The scales of the in-flows are not great enough to 
produce new black holes, but are consistent with the 
re-fuelling model.
However,
the simulations only predict in-flow to within a few hundred pc of the
galaxies' cores, whereas typical quasar accretion disks 
extend out to only pc scales \cite{pe97}. It is not clear whether
dissipation is sufficient to allow the gas to fall further.

The formation of binary quasars is qualitatively consistent with the
reactivation process, provided that the progenitors (or their central black
holes) are sufficiently massive. 
The observational result that the activation radius is between
50 kpc and 100 kpc (Sec.~\ref{section:orbits}) implies that
quasar formation does not occur until quite late in the 
collision process. This is consistent with the $N$-body simulations 
of Barnes \& Hernquist \shortcite{ba96}, which showed that gas inflows
took some time to occur.

The separation scale of the binaries,
together with the requirement of a massive
black hole, is also suggestive of 
the fact that quasars form in larger galaxies.
If this is the case, encounters between dwarfs and large galaxies 
can result in the formation of a quasar, but only collisions between
two large galaxies can result in a binary quasar. 
The fraction of merger-related quasars which are binaries is 
comparable to
the fraction of collisions in which both galaxies have central black holes 
of $\ga 10^6 M_{\sun}$.
Assuming a Schechter \shortcite{sc76} mass function for the
galaxies, and that $\sim 20$ per cent of those heavier than 
$M_* \simeq 10^{11} M_{\sun}$
contain large black holes \cite{ma98}, about 1 per cent of 
all merger-formed quasars should be in binary systems\footnote{It
also implies that only $\sim 1$ per cent of multiple quasars 
should be triples etc., justifying the assumption that all observed
triples and quadruples are actually lenses.}. 
Combined with the observation that only 1 quasar in $\sim 1000$
is a binary, this implies that about 10 per cent of quasar activity is 
related to galactic interactions. 
This is probably an underestimate of the fraction, however, 
as interactions between two large galaxies are almost 
certainly more efficient than collisions involving a dwarf galaxy.
Further, if quasars all have a dusty torus which is both optically 
thick and subtends a large fraction of the solid angle 
around the central engine, both the quasars in a pair must be close
to face-on for it to be observed as a binary.
Extrapolating
from the observation that $\sim 25$ per cent
of all Seyferts are of Type I
\cite{pe97}, 1 in $\sim4$ of all quasars are visible, but
only 1 in $\sim 16$ binaries are detected as such.
It is thus possible that $\sim 50$ per cent
of all quasars are formed during
galactic interactions and collisions.

\subsubsection{Evolution}
\label{section:df}

The theoretical and observational evidence discussed in 
Sec.~\ref{section:activation} is consistent with binary quasars
forming during galactic mergers. 
Once the merger has begun to settle, the quasars can be 
treated as point-masses moving in a static, isothermal
halo.
Their orbits slowly decay, as they lose energy and 
angular momentum through dynamical friction 
(e.g.\ Binney \& Tremaine 1987), eventually either merging
(e.g.\ Makino \& Ebisuzaki 1994) or forming a hard
binary that lasts for a Hubble time or more
(e.g.\ Begelman, Blandford \& Rees 1980; Rajagopal \& Romani 1996).
If the quasars have mass $M \simeq 10^9 M_{\sun}$ by this stage,
and are in circular orbits of speed $\sqrt{2} \sigma$ (where
$\sigma$ is the velocity dispersion of the halo), the time 
spent with separation at radius $r$ is \cite{bi87}
\begin{equation}
\left| \frac{{\rmn d} t}{{\rmn d} r}\right| \simeq 3.3
\frac{\sigma}{G M \ln (\Lambda)} r,
\label{equation:dtdr_df}
\end{equation}
where $\ln (\Lambda) \simeq \ln (2 r_* \sigma^2 / G M)$ is the Coulomb
logarithm, which
characterises the strength of the interaction, and 
$r_* \simeq 10$ kpc \cite{la93}.
Hence
${\rmn d} p/{\rmn d}r \propto r$, as used in Sec.~\ref{section:circles}
to fit the observed separation distribution. The orbital decay
proceeds at a greater rate as $r$ decreases, potentially
explaining the small separation hole (Sec.~\ref{section:orbits}).

The time-scale for the decay, $t_{\rmn DF}$, is found 
by integrating \eq{dtdr_df} from $r_{\rmn max}$ to 0.
Following Binney \& Tremaine \shortcite{bi87},
\begin{eqnarray}
t_{\rmn DF}
& \simeq & \frac{1.65}{\ln (\Lambda)}
\frac{r_{\rmn max}^2 \sigma}{G M} \nonumber \\
& \simeq &
4 \times 10^{9}
\left(\frac{r_{\rmn max}}{{\rm 20\,\, kpc}}\right)^2
\frac{\sigma}{{\rmn 200\,\, km\,\, s}^{-1}}
\frac{10^9 M_{\sun}}{M} \,\,{\rm yr},
\label{equation:t_df}
\end{eqnarray}
where the scale values have been chosen to give the {\em shortest}
plausible decay time\footnote{The value of $r_{\rmn max}$ used is half the
activation radius inferred in Sec.~\ref{section:orbits}, 
as it is possible that {\em both} quasars orbit the centre of the 
halo, whence their distance between them is twice their orbital radius.
Also, the black holes themselves might start with $M \ll 10^9 M_{\sun}$,
but they must be associated with most of their eventual mass; hence
the choice of the canonical mass.}.
The dynamical time-scale of the binaries is then close to a 
Hubble time, and much longer than both the `settling' time
of the merged halo ($< 10^9$ yr; Barnes 1992) and the 
expected quasar lifetime. A black hole of initial mass $\sim 10^7
M_{\sun}$ (as implied by dynamical measurements of local 
galaxies; Kormendy et al.\ 1996) accreting at 
the Eddington limit \cite{pe97} would reach $\sim 10^{10}
M_{\sun}$ in much less than $10^9$ yr. No such massive black holes 
are observed now (e.g.\ Kormendy \& Richstone 1995), so 
this places an upper limit on quasar
masses and hence their lifetimes. Comparison of the three time-scales 
implies that binary quasars are short-lived, 
existing only whilst the host galaxies are actively merging.
The most likely explanation for the death of the binaries 
(and other quasars in mergers) 
is that the in-flow of gas ceases,
probably as the merger becomes more stable.
Moreover, by the time the orbits decay due to dynamical friction,
the merging black holes are already quiescent, and the
observed binary separation distribution is a random sampling of the chaotic
phase of the merger.

If the above model is correct, the hosts of binary quasars should
be distinct (in the case of a `new-born' binary), or,
more likely, highly distorted.
The host galaxies of 
MGC~2214+3350 A and B \cite{mu98} have been detected by
the {\em HST}, and in fact out-shine the quasars 
in the $H$-band \cite{ko99b}. The galaxies show no obvious signs of 
being distorted, which implies both that their physical separation
is considerably greater than their projected separation of $\sim 20$ kpc,
and that the nuclei have only recently become active.
Comparable observations -- which need not be prohibitively deep --
of some of the other ambiguous pairs ought to prove similarly revealing.

\section{Conclusions}
\label{section:conclusion}

Djorgovski \shortcite{dj91} and Kochanek et al.\ \shortcite{ko99a}
have both argued that most of the known wide separation quasars pairs
are not the result of gravitational
lensing, as summarised in Sec.~\ref{section:pairs}.
However, there have also been some valid objections
to the interpretation of the pairs as physical binaries.
The strongest was that 
the spectra of the pairs appear to be very similar.
Sec.~\ref{section:spectra} consists of a quantitative evaluation
of this claim for the three quasar pairs discovered as part of 
the LBQS, and shows that, despite appearances, none of 
these pairs can be rejected as binaries on spectral grounds.
Extrapolating to the rest of the ambiguous quasar pairs,
the simplest conclusion is that they are all binary quasars.

The adoption of the binary hypothesis for the entire population
of 16 pairs listed in Table~\ref{table:binaries} provides a 
significant data-set of binary quasars, and even if some
do turn out to be lenses, the statistical properties of the population
will not be greatly changed.
Assuming quasars are the nuclei of large galaxies,
the velocity differences of the pairs are consistent with them
belonging to bound systems, which supports the idea that they
are part of galactic mergers.
Both the orbital and deprojection analyses presented in 
Sec.~\ref{section:orbits} suggest that 
the galactic nuclei become active at separations of
between 50 kpc and 100 kpc,
depending on the cosmology and the ellipticity of the quasars'
orbits. This is certainly consistent with the scales at which
the tides from large galaxies become important.
However the time-scales suggested for the orbits of the nuclei
to decay are very large -- certainly longer than the lifetime
of a quasar accreting at the Eddington limit.
Indeed it seems probable that the quasars 
remain active only whilst the host galaxies are merging,
in which case most of the binaries should exist in very
distorted hosts, and not in a relaxed merger. 
The observation that the host galaxies of MGC~2214+3550 A and B
\cite{ko99b} 
are undisturbed then suggests that this is a 
very young binary.

Whilst it is clear that the physics of merging galaxies
and active galactic nuclei are far from solved, there are
also a number of questions that can be answered observationally.
Further spectroscopy and radio observations of a number of 
the quasar pairs should reveal whether they are binaries\footnote{During 
the preparation of this paper, Q~1216+5032 \cite{ko99a}
was confirmed as a binary,
and RXJ~0911+0551 \cite{ba97} was found to be a lens.},
although there are also several which are still not classified 
despite major observational efforts.
The nature of the `turn on' radius will be
more strongly constrained by the 2 degree 
Field quasar survey \cite{bo98},
which will include analysis of the nearby companions of $\sim 3 \times 10^4$
quasars. On a longer time-scale, the Sloan Digital Sky Survey
\cite{gu95} will observe $\sim 10^5$ quasars, and is expected to 
find several hundred lenses and pairs, an order of magnitude 
increase on the number known presently. Both the lower limits on
mass of the quasars' host galaxies
and the constraints on the activation radius
could be determined to within $\sim 10$ per cent from such a sample. 

\section*{Acknowledgments}

Paul Hewett and Craig Foltz kindly supplied the spectra of the
LBQS quasar pairs.
The discussion of the evolution and fate of the binary quasars was
enhanced by
stimulating discussions with Chris Kochanek (who also suggested 
a number of improvements as the referee), John Kormendy, Paul Nulsen
and Matthew O'Dowd.
Extensive use was made of the Center for Astrophysics-Arizona 
{\em Space Telescope} Lens (CASTLe)
survey world-wide web site 
({\tt http:/$\!$/cfa-www.harvard.edu/glensdata}), maintained by
Chris Kochanek, Emilio Falco, Chris Impey, Joseph Leh\'{a}r, 
Brian McLeod and Hans-Walter Rix. 
DJM was supported by an Australian Postgraduate Award.

{}

\bsp
\label{lastpage}
\end{document}